\documentclass[sigconf,nonacm]{acmart}

\setcopyright{none}

\renewcommand\footnotetextcopyrightpermission[1]{}

\usepackage{graphicx}
\usepackage{booktabs}
\usepackage{algorithm}
\usepackage{algpseudocode}
\usepackage{array}

\begin{document}

\title{Novelty-Aware Agentic Retrieval: Comparing Research
Contributions Through Structured Multi-Step Reasoning}

\author{Shou-Tzu Han}
\affiliation{%
  \institution{Department of Computer Science}
  \institution{University of South Dakota}
  \city{Vermillion}
  \state{South Dakota}
  \country{USA}}
\email{shoutzu.han@coyotes.usd.edu}

\begin{abstract}
Scientific literature search is an information retrieval (IR) task in which
ranked lists are insufficient: a researcher entering a new area needs to know
not only \emph{which} papers are relevant, but how they relate: where they
overlap, how they differ, and what problem--method combinations are absent.
Standard retrieval-augmented generation (RAG) summarizes documents
independently, discarding exactly this comparative signal. We present the
\emph{Novelty-Aware Research Agent}, a prototype agentic retrieval system that
layers structured multi-step reasoning on a RAG pipeline through six
typed-contract components: query analysis, a ReAct-style retrieval loop,
relevance ranking, schema-guided contribution extraction, a three-pass
comparison agent, and answer generation. Beyond returning relevant papers, it
produces structured comparison artifacts: per-paper contribution records,
paper-level overlaps, and a problem~$\times$~method gap matrix.
On a 100-paper corpus, our central result is that the system supports five
structured comparison capabilities that a standard RAG baseline supports none
of, while remaining query-sensitive. Across the three main queries, no paper
appears in all three top-5 sets and the mean pairwise Jaccard similarity is
0.12; in an extended seven-query evaluation, the same pattern holds across ten
total queries, with mean pairwise Jaccard similarity of 0.115 and 18 of 29
distinct retrieved papers appearing in only one query. Under author-assigned
graded relevance, the ranker attains mean Precision@5 of 1.000 and nDCG@5 of
0.752 on the three main queries, ahead of BM25, dense, and hybrid retrieval;
over all ten queries, Precision@5 remains high but non-saturated at 0.980, with
nDCG@5 of 0.739. Schema compliance is 86.7\% on the main queries and 84.0\% over
the ten-query set, and a validation of 20 sampled empty gap-matrix cells yields
gap precision of 0.600. We discuss the latency--structure trade-off
inherent to agentic retrieval and identify corpus scale, author-assigned
relevance labels, and limited independent evaluation as the primary limitations
of the prototype.
\end{abstract}



\maketitle

\section{Introduction}

The integration of large language model (LLM) agents into information retrieval
has reshaped how users find and consume information. Agentic retrieval
systems, those combining tool use, memory, reasoning, and planning, can
decompose queries, retrieve evidence, and synthesize responses in ways that go
beyond returning a ranked list of documents. Yet for one common and important IR
task, scientific literature search, even agentic systems tend to fall back on a
familiar pattern: retrieve relevant documents, then summarize each one
independently.

This pattern is inadequate for the underlying user need. A researcher entering a
new area does not primarily want a list of papers or a set of disconnected
summaries. They want to understand the \emph{structure} of a body of work: which
papers address the same problem, which propose genuinely different methods, and
crucially, what problem--method combinations have \emph{not} yet been explored.
These are comparative and corpus-level questions. A retrieval system that
summarizes documents independently discards exactly the signal needed to answer
them. Concretely, the payoff of answering them is that a researcher can see, in
one structured view, which retrieved papers converge and which problem--method
combinations remain open, rather than reconstructing that map by reading and
cross-referencing a ranked list by hand.

We frame this as a novelty-aware retrieval problem and present the
\emph{Novelty-Aware Research Agent}, a prototype agentic retrieval system that
layers structured comparison reasoning on top of a retrieval-augmented
generation (RAG) pipeline. Rather than asking ``what does this paper say?'' the
system asks ``what does this retrieved set collectively cover, where does it
converge, and what is absent?''

The system retrieves papers from a domain corpus, ranks them by query relevance,
extracts a structured contribution record for each, and then runs a dedicated
comparison stage that identifies overlaps, differentiating aspects, and a
problem~$\times$~method gap matrix. The final output is not a single prose
summary but a set of typed, auditable artifacts that a user (or a downstream
system) can inspect and act on.

\medskip
\noindent\textbf{Contributions.}
\begin{enumerate}
  \item A six-component agentic retrieval pipeline with typed inter-component
        contracts, combining a ReAct-style retrieval loop, relevance ranking,
        schema-guided extraction, and a multi-pass comparison stage.
  \item A three-pass Comparison Agent that operates on structured contribution
        records rather than raw retrieved text, producing paper-level overlaps,
        per-paper differentiation, and a deterministic problem~$\times$~method
        gap matrix.
  \item An evaluation on a 100-paper agentic AI and retrieval corpus, including
        query-sensitivity analysis, graded-relevance retrieval metrics
        (Precision@5, nDCG@5, Recall@5, MRR), schema-compliance failure
        analysis, a ReAct sparse-query stress test, deterministic gap-matrix
        construction, a baseline comparison, BM25/dense/hybrid retriever
        comparison, component ablation, deterministic gap validation, gap error
        analysis, a qualitative case study, an indicative single-rater
        external usefulness assessment collected on an earlier system version,
        and an extended seven-query evaluation.
  \item A working open-source prototype implementation with a web interface that streams
        the six-stage retrieval process in real time.\footnote{Code:
        \texttt{https://github.com/52147/Novelty-Aware-Research-Agent} \quad
        Live demo:
        \texttt{https://debrah1-novelty-aware-research-agent.hf.space}}
\end{enumerate}

We position this as a prototype system paper. The contribution is the retrieval
architecture and the comparison artifacts it produces, not a large-scale
benchmark; we are explicit throughout about the scale and evaluation limitations
of the current corpus.

\section{Related Work}

\textbf{Agentic IR and RAG.}
LLM agents extend retrieval beyond single-shot ranking through query
decomposition, iterative search, and tool use. Multi-agent frameworks such as
AutoGen~\cite{wu2023autogen} and MetaGPT~\cite{hong2023metagpt} coordinate agents
for complex tasks, AgentBench~\cite{liu2023agentbench} evaluates agent behavior
across environments, and a recent survey~\cite{wang2023survey} reviews agent
architectures. These establish the agentic paradigm but do not target the IR task
of comparing retrieved documents at the level of their contributions. Our system
builds on retrieval-augmented generation (RAG)~\cite{lewis2020rag}, including its
adaptive and self-critical variants Self-RAG~\cite{asai2023selfrag} and corrective
retrieval~\cite{yan2024crag}, using FAISS exact nearest-neighbor search as a
substrate. It departs from typical RAG output: where standard RAG concatenates
passages into one response, our system retains each retrieved paper as a distinct
unit, extracts a structured record, and reasons across records.

\textbf{Reformulation, structured extraction, and comparison.}
Effective retrieval often depends on reformulating the query.
ReAct~\cite{yao2022react} interleaves reasoning and action in a
Thought~$\rightarrow$~Action~$\rightarrow$~Observation loop; we apply this at the
retrieval stage, placing reformulation under explicit agent control rather than a
single fixed embedding. JSON schema-guided decoding~\cite{willard2023efficient}
constrains generation to structurally valid output; we use it to extract a fixed
four-field contribution record per paper, so downstream comparison operates over
uniform inputs. Chain-of-thought prompting~\cite{wei2022cot} and deployment
platforms such as OpenAgents~\cite{xie2023openagents} are adjacent reasoning and
agent-deployment work covered by our corpus. We are not aware of prior agentic
retrieval systems that combine structured contribution extraction, cross-paper
comparison, and deterministic problem--method gap-matrix construction in a single
literature-comparison pipeline. Table~\ref{tab:related-work} positions our system
relative to representative prior work across five capability dimensions.

\textbf{Automated literature review and gap analysis.}
A parallel line of work automates systematic literature reviews (SLRs) with LLM
agents. Sami et al.~\cite{sami2024slr} propose a multi-agent system that
automates the full SLR workflow, including agents that retrieve, filter, and
summarize papers and surface trends and gaps, and Moses et
al.~\cite{moses2025gapaware} introduce a gap-aware agentic workflow that combines
structured synthesis, knowledge-graph modeling, and perspective-guided
questioning to identify gaps in coverage, reasoning, or evidence through graph
traversal and contrastive retrieval. Our system shares the multi-agent and
gap-surfacing motivation but differs in target and mechanism: rather than
generating a written review narrative, it treats the retrieved set as structured
data and constructs a deterministic problem~$\times$~method gap matrix over typed
contribution records, so that the gap decision is a reproducible structural
computation rather than generated text.

\begin{table*}[t]
\caption{Positioning of the proposed system relative to representative prior work.}
\label{tab:related-work}
\centering
\resizebox{\textwidth}{!}{%
\begin{tabular}{lccccc}
\toprule
\textbf{System / Work} & \textbf{RAG} & \textbf{Agent Loop}
  & \textbf{Structured Extraction} & \textbf{Cross-Paper Comparison}
  & \textbf{Problem--Method Gap Matrix} \\
\midrule
RAG \cite{lewis2020rag}              & Yes       & No  & No  & No  & No \\
ReAct \cite{yao2022react}            & Tool-dep. & Yes & No  & No  & No \\
Self-RAG \cite{asai2023selfrag}      & Yes       & Yes & No  & No  & No \\
AutoGen \cite{wu2023autogen}         & No        & Yes & No  & No  & No \\
MetaGPT \cite{hong2023metagpt}       & No        & Yes & No  & No  & No \\
AgentBench \cite{liu2023agentbench}  & No        & No  & No  & No  & No \\
SLR multi-agent systems \cite{sami2024slr}   & Yes & Yes & Partial & Yes & No \\
Gap-aware workflows \cite{moses2025gapaware} & Yes & Yes & Yes     & Yes & Partial \\
\textbf{Novelty-Aware Research Agent (ours)} & Yes & Yes & Yes & Yes & Yes \\
\bottomrule
\end{tabular}%
}
\end{table*}

\section{System Architecture}

\begin{figure*}[t]
  \centering
  \includegraphics[width=0.95\textwidth]{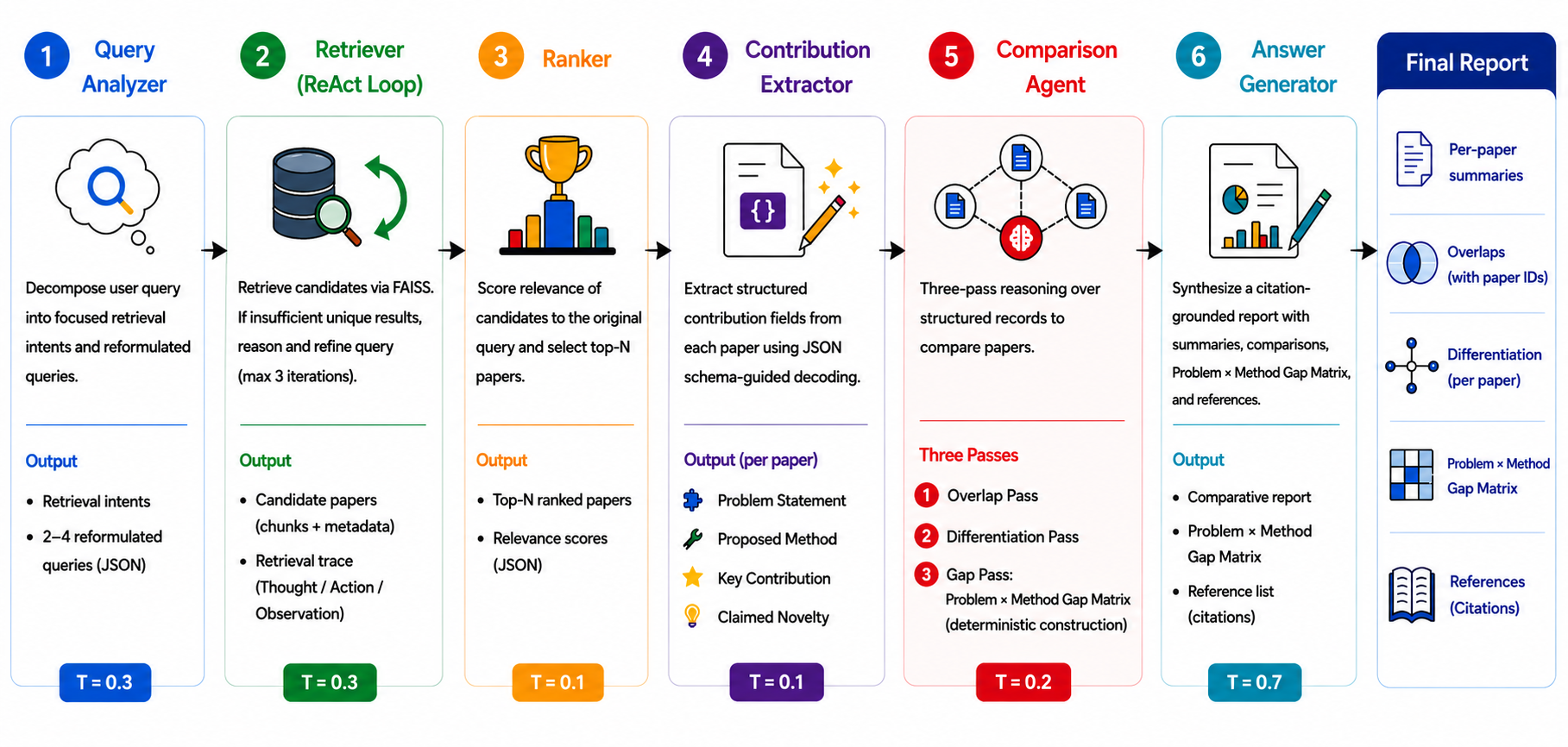}
  \caption{End-to-end agentic retrieval pipeline. A user query is decomposed and
  reformulated, candidate papers are retrieved and ranked, structured
  contribution records are extracted under a schema, three comparison passes
  produce overlaps / differentiation / a gap matrix, and a final
  citation-grounded report is generated.}
  \label{fig:pipeline}
\end{figure*}

The system comprises six components arranged in a sequential agentic retrieval
pipeline with defined input/output contracts. Figure~\ref{fig:pipeline} shows
the data flow and Algorithm~\ref{alg:pipeline} formalizes the procedure.
Lines~3--10 implement the ReAct retrieval loop; lines~13--15 implement the three
comparison passes.

\begin{algorithm}[t]
\caption{Novelty-Aware Agentic Retrieval}
\label{alg:pipeline}
\begin{algorithmic}[1]
\Require query $q$, corpus $C$, min-papers $k$, top-$n$ $N$
\Ensure structured comparison report $R$
\State $Q \gets \textsc{QueryAnalyzer}(q)$ \Comment{$T{=}0.3$}
\State $iter \gets 0$
\While{$iter < 3$}
    \State $P \gets \textsc{FAISS.search}(Q.queries[iter], C)$
    \If{$|\textsc{unique}(P)| \geq k$}
        \State \textbf{break} \Comment{ReAct STOP}
    \EndIf
    \State $q' \gets \textsc{LLM.reason}(q, P)$ \Comment{ReAct REFINE}
    \State $Q.queries.\textsc{append}(q')$;\ $iter{+}{+}$
\EndWhile
\State $P_n \gets \textsc{Ranker}(P, q, N)$ \Comment{$T{=}0.1$}
\State $E \gets \textsc{Extractor}(P_n)$ \Comment{schema;\ $T{=}0.1$}
\State $O \gets \textsc{OverlapPass}(E)$ \Comment{$T{=}0.2$}
\State $D \gets \textsc{DifferentiationPass}(E)$ \Comment{$T{=}0.2$}
\State $G \gets \textsc{GapMatrix}(E)$ \Comment{deterministic}
\State $R \gets \textsc{AnswerGenerator}(q,E,O,D,G)$ \Comment{$T{=}0.7$}
\State \Return $R$
\end{algorithmic}
\end{algorithm}

\subsection{Query Analyzer}
The entry point decomposes the user query into focused retrieval intents and
2--4 reformulated queries, operating at temperature $T{=}0.3$ under JSON schema
constraints. Reformulation is included to support sparse or underspecified
retrieval settings, with a modest measurable benefit on the present 100-paper
corpus (Section~\ref{sec:results}).

\subsection{Retriever (ReAct Loop)}
The retriever performs FAISS IndexFlatL2 search over sentence-transformer
embeddings with ReAct-style refinement. If fewer than three unique papers meet
an $\mathrm{L2} < 400$ distance threshold, the agent emits a Thought explaining
the shortfall and an Action query reformulation, retrying up to three
iterations. In the main broad-query experiments retrieval succeeded in one
iteration on every run. We therefore additionally conduct a sparse-query stress
test (Section~\ref{sec:react}), in which the loop activates on three of five
strict-threshold queries and recovers sufficient coverage in one of the three
triggered cases.

\subsection{Ranker}
The ranker scores each retrieved candidate for query relevance at $T{=}0.1$ and
returns the top-$N$. Low temperature yields consistent, near-deterministic
scoring. As Section~\ref{sec:results} shows, the ranker selects substantially
different paper sets across queries, evidence that scoring is query-sensitive
rather than returning a fixed corpus ordering.

\subsection{Contribution Extractor}
For each top-ranked paper, schema-guided decoding enforces a four-field record:
\textit{Problem Statement}, \textit{Proposed Method}, \textit{Key Contribution},
and \textit{Claimed Novelty}, at $T{=}0.1$. We define schema compliance as the
fraction of papers for which all four fields are non-empty and
non-placeholder. Structural validity is guaranteed by decoding; field
completeness depends on the input paper, which is why compliance can fall below
100\% for atypically structured papers.

\subsection{Comparison Agent}
The comparison agent is the core contribution and operates on structured records
rather than raw retrieved text, which bounds the reasoning space and makes
outputs auditable. It runs three sequential passes:
\begin{itemize}
  \item \textbf{Overlap Pass.} Identifies papers sharing the same problem
        formulation, dataset, or method family, returning paper identifiers with
        a shared-element label.
  \item \textbf{Differentiation Pass.} Identifies what each paper does
        distinctly in method, scope, or claimed contribution.
  \item \textbf{Gap Pass.} Maps retrieved paper identifiers to canonical problem
        and method labels using a fixed taxonomy; the final
        problem~$\times$~method matrix is constructed deterministically, and
        empty cells are reported as candidate corpus-level gaps within the
        retrieved paper slice (Section~\ref{sec:gapmatrix}).
\end{itemize}

\subsection{Answer Generator}
The final stage synthesizes a report at $T{=}0.7$, producing per-paper
summaries, a synthesis paragraph, and a citation-grounded reference list.
Higher temperature here favors readable natural-language synthesis, in contrast
to the precision-oriented extraction and ranking stages.

\section{Implementation}

\subsection{Technology Stack}
The system is implemented in Python~3.9 using the OpenAI API~\cite{openai2023gpt4},
with GPT-4o as the primary backbone for the main experiments and GPT-4o-mini and
GPT-4.1 used in the cross-model robustness analysis (Section~\ref{sec:multimodel}).
The system uses FAISS IndexFlatL2 for exact nearest-neighbor retrieval,
Sentence-Transformers (all-MiniLM-L6-v2, 384-dimensional embeddings) for
encoding, and Pydantic~v2 for typed inter-component contracts. A FastAPI backend
streams the six-stage retrieval process to a web interface using server-sent
events, allowing users to observe each stage complete in real time.

\subsection{Corpus}
The evaluation corpus contains 100 papers spanning the agentic AI and retrieval
domain: multi-agent frameworks, reasoning techniques, tool-use systems, agent
deployment platforms, evaluation and survey work, and a substantial set of
retrieval-augmented generation papers. Representative papers include
Self-RAG~\cite{asai2023selfrag}, Reflexion~\cite{shinn2023reflexion},
CAMEL~\cite{li2023camel}, Generative Agents~\cite{park2023generative},
AgentVerse~\cite{chen2023agentverse}, ReWOO~\cite{xu2023rewoo}, and
MetaAgents~\cite{li2023metaagents}. Each
paper is chunked into abstract, introduction, and conclusion sections, producing
300 FAISS vectors. The corpus defines the system's entire search space; the
system does not retrieve from the open web.

\subsection{Temperature Configuration}
Temperature is assigned per component following standard practice: low values
for precision-critical stages (ranking and extraction at $T{=}0.1$, comparison
at $T{=}0.2$), moderate values for query analysis and ReAct reasoning
($T{=}0.3$), deterministic construction for the gap matrix, and a higher value
($T{=}0.7$) for natural-language synthesis.

\section{Experiments and Results}
\label{sec:results}

We evaluate using (a) automated retrieval and compliance metrics, (b)
graded-relevance retrieval quality (Precision@5, nDCG@5, Recall@5, MRR), (c)
schema failure analysis, (d) query-sensitivity analysis, (e) a ReAct
sparse-query stress test, (f) deterministic gap-matrix construction, (g) a
baseline comparison, (h) BM25/dense/hybrid retriever comparison,
(i) component ablation, (j) deterministic gap validation,
(k) gap error analysis, (l) a qualitative case study, (m) an indicative single-rater
external usefulness assessment collected on an earlier system version, and
(n) an extended seven-query evaluation. Three main runs were executed with
distinct broad comparison queries on the expanded 100-paper corpus:
\begin{itemize}
  \item \textbf{R1:} ``Compare multi-agent LLM frameworks for collaborative
        reasoning''
  \item \textbf{R2:} ``What evaluation methods exist for LLM reasoning agents?''
  \item \textbf{R3:} ``Compare verbal reinforcement and role-playing approaches
        in LLM agents''
\end{itemize}
Unless otherwise stated, the main tables report these three primary runs. We
additionally report an extended seven-query evaluation (R4--R10) in
Section~\ref{sec:extended} to test whether the retrieval behavior remains stable
across a wider query set.

\subsection{Automated Metrics}
Table~\ref{tab:metrics} reports automated metrics. Retrieval returns nine
to ten candidates per query in these runs; the ranker reduces them to a top-5 set.
Schema compliance averages 86.7\%, and the comparison agent produces 2--3 overlaps
and 4--5 report-level gaps per run. End-to-end latency averages 23 seconds.

\begin{table}[t]
\caption{Automated metrics across three runs (100-paper corpus).}
\label{tab:metrics}
\begin{tabular}{lrrrr}
\toprule
\textbf{Metric} & \textbf{R1} & \textbf{R2} & \textbf{R3} & \textbf{Avg} \\
\midrule
Corpus size          & 100  & 100  & 100   & 100  \\
Candidates retrieved & 9    & 10   & 9     & 9.3  \\
ReAct iterations     & 1    & 1    & 1     & 1.0  \\
Schema compliance    & 80\% & 80\% & 100\% & 86.7\% \\
Overlaps detected    & 3    & 3    & 2     & 2.7  \\
Differences          & 5    & 5    & 5     & 5.0  \\
Gaps identified      & 5    & 4    & 5     & 4.7  \\
Runtime (s)          & 22.3 & 23.5 & 22.9  & 22.9 \\
\bottomrule
\end{tabular}
\end{table}

\subsection{Query-Sensitive Retrieval}
The central retrieval result is that the ranker selects substantially different
paper sets across queries (Table~\ref{tab:top5}). No paper appears in all three
top-5 result sets. Across the 15 ranked slots the system retrieves 12 distinct
papers, 9 of which (75\%) are query-exclusive, appearing in only a single run.
The mean pairwise Jaccard similarity across the three top-5 sets is 0.12,
indicating the retrieved sets are roughly 88\% distinct on average; runs R2 and
R3 share no papers at all. This indicates the ranking stage responds to query
semantics rather than returning a fixed corpus ordering, a necessary property
for an agentic retrieval system whose value depends on tailoring the retrieved
set to the user's specific comparative question. Notably, several papers added in
the corpus expansion (OpenAgents, ART~\cite{paranjape2023art}, and
RAP~\cite{hao2023rap}) are selected into top-5 sets, confirming that the enlarged
corpus actively changes retrieval rather than being ignored. The query-sensitivity statistics indicate that the enlarged corpus does not
collapse the system into a fixed retrieval pattern.

\begin{table}[t]
\caption{Top-5 papers selected per run on the 100-paper corpus. No paper appears
in all three runs.}
\label{tab:top5}
\begin{tabular}{lll}
\toprule
\textbf{R1: Multi-agent} & \textbf{R2: Evaluation} & \textbf{R3: Verbal RL} \\
\midrule
AgentVerse   & AgentBench       & Reflexion       \\
AutoGen      & AgentSurvey      & CAMEL           \\
AgentSurvey  & ART              & AgentVerse      \\
AgentBench   & Chain-of-Thought & Gen.~Agents     \\
OpenAgents   & RAP              & Inner Monologue \\
\bottomrule
\end{tabular}
\end{table}

\subsection{Retrieval Quality}
\label{sec:ir-metrics}
Query-sensitivity shows the retrieved sets \emph{differ} across queries, but not
whether they are \emph{good}. To assess ranking quality directly, we labeled the
relevance of corpus papers for each of the three main queries on a four-point
graded scale (3~=~highly relevant, 2~=~relevant, 1~=~marginal, 0~=~not relevant)
and computed Precision@5, nDCG@5, Recall@5, and mean reciprocal rank (MRR).
Relevance labels are author-assigned; we report this as a limitation in
Section~\ref{sec:limitations}, and treat the labels as a small-scale,
single-annotator gold standard rather than a benchmark-grade resource.

Table~\ref{tab:ir-metrics} reports the results. Precision@5 is 1.0 on all three
queries: every paper the ranker placed in the top-5 was judged at least
marginally relevant. We read this cautiously: perfect precision on
author-assigned labels reflects that the ranker avoids clearly off-topic papers
rather than that retrieval is solved. The more informative signals are nDCG@5 and
Recall@5. Mean nDCG@5 is 0.752, with the evaluation query (R2) lowest at 0.588:
although all five retrieved papers were relevant, their \emph{ordering} did not
match the ideal graded ranking, indicating the ranker captures relevance better
than fine-grained priority. Mean Recall@5 is 0.527, meaning a top-5 cutoff
recovers roughly half of the papers labeled relevant in the corpus, expected
given that several queries have more than five relevant papers, and a useful
characterization of the precision--recall trade-off at this cutoff. MRR is 1.0
throughout: the top-ranked paper was always relevant.

\begin{table}[t]
\caption{Retrieval quality on the three main queries (author-assigned graded
relevance). Precision@5 and MRR are saturated; nDCG@5 and Recall@5 are the more
discriminative measures.}
\label{tab:ir-metrics}
\begin{tabular}{lrrrr}
\toprule
\textbf{Query} & \textbf{P@5} & \textbf{nDCG@5} & \textbf{Recall@5} & \textbf{MRR} \\
\midrule
R1 (multi-agent) & 1.000 & 0.719 & 0.455 & 1.000 \\
R2 (evaluation)  & 1.000 & 0.588 & 0.500 & 1.000 \\
R3 (verbal RL)   & 1.000 & 0.950 & 0.625 & 1.000 \\
\midrule
\textbf{Mean}    & 1.000 & 0.752 & 0.527 & 1.000 \\
\bottomrule
\end{tabular}
\end{table}

\subsection{Retriever Comparison}
To compare the ranker against simpler retrieval alternatives, we evaluated BM25,
dense retrieval, hybrid retrieval, and the full ranker using the same
author-assigned graded relevance labels. Table~\ref{tab:retriever-comparison}
reports mean performance across the three main queries.

The full ranker achieves the highest Precision@5 and nDCG@5, improving mean
Precision@5 to 1.000 (versus 0.733 for BM25 and 0.667 for both dense and hybrid
retrieval) and improving mean nDCG@5 to 0.752 (versus 0.640 for BM25, 0.595 for
dense retrieval, and 0.685 for hybrid retrieval). On the larger corpus BM25 is a
stronger baseline than before, narrowing but not closing the precision gap.
Because the full ranker outputs five selected papers, we compare it to the other
retrievers primarily at @5; @10 metrics are not defined for the capped
full-ranker output.

\begin{table}[t]
\caption{Retriever comparison across BM25, dense retrieval, hybrid retrieval,
and the full ranker. Metrics are averaged over the three main queries using
author-assigned graded relevance.}
\label{tab:retriever-comparison}
\centering
\resizebox{\columnwidth}{!}{%
\begin{tabular}{lrrrrrr}
\toprule
\textbf{System} & \textbf{P@5} & \textbf{P@10} & \textbf{nDCG@5} & \textbf{nDCG@10} & \textbf{R@5} & \textbf{R@10} \\
\midrule
BM25 & 0.733 & 0.433 & 0.640 & 0.587 & 0.396 & 0.457 \\
Dense & 0.667 & 0.333 & 0.595 & 0.525 & 0.343 & 0.343 \\
Hybrid & 0.667 & 0.500 & 0.685 & 0.658 & 0.343 & 0.520 \\
Full ranker & 1.000 & -- & 0.752 & -- & 0.527 & -- \\
\bottomrule
\end{tabular}%
}
\end{table}

\subsection{Reformulation Ablation}
We tested whether the number of reformulated queries affects retrieval by
re-running each main query with one, two, and four reformulations and measuring
Recall@5. Mean Recall@5 was 0.493 with one reformulation, 0.527 with two, and
0.493 with four, a small benefit at two reformulations driven by the evaluation
query (R2: 0.400 to 0.500), with the other two queries flat. The effect is modest:
on a focused corpus the base query already retrieves much of the relevant
neighborhood, so additional reformulations add little. We retain reformulation as
an architectural feature whose value is more likely to appear at larger corpus
scales and under the sparse-retrieval conditions where the ReAct loop activates
(Section~\ref{sec:react}).

\subsection{Schema Compliance and Failure Analysis}
Two runs achieved 80\% compliance (4/5) and one achieved 100\% (5/5), for a mean of 86.7\%. Failures occurred on papers with
survey-like structure, merged contribution and novelty fields, or broader
framework papers whose contribution statements did not map cleanly to the
four-field schema. Structural validity of the JSON is
always guaranteed by constrained decoding; the failures are field-completeness
failures on atypically structured papers. The recommended fix is a fallback
extraction prompt targeting such papers.

\subsection{ReAct Loop Activation Under Sparse Retrieval}
\label{sec:react}
In the main broad-query experiments the ReAct refinement loop did not activate,
because the 100-paper corpus returns sufficient candidates for any broad query on
the first iteration. To test the loop directly, we constructed a
sparse-retrieval stress test: five narrow queries evaluated under a strict
relevance threshold that forces first-pass retrieval below the minimum-papers
requirement.

Table~\ref{tab:react} reports the result. Under the default threshold the loop
remains dormant on all five queries. Under the strict threshold the loop
activates on three of five queries, issuing one to two refinement iterations, and
recovers to sufficient coverage ($\geq 3$ papers) on one of the three
activations; for example, increasing retrieval from one paper to six for the
paged-memory query. This demonstrates that the refinement mechanism is
functional and beneficial when retrieval is sparse, while remaining correctly
inactive when the corpus already supplies enough candidates.

\begin{table}[t]
\caption{ReAct loop activation under a sparse-retrieval stress test. The loop
fires on 3/5 queries and recovers coverage on 1/3 activations. ``Iters'' counts
total iterations (initial pass plus refinements), so the number of refinement
steps is $\text{Iters}-1$.}
\label{tab:react}
\begin{tabular}{p{0.40\columnwidth}rcrr}
\toprule
\textbf{Query (sparse)} & \textbf{Iter-1} & \textbf{Refine} & \textbf{Iters} & \textbf{Final} \\
\midrule
verbal self-reflection memory   & 3 & no  & 1 & 3 \\
chunked cross-attention retr.    & 6 & no  & 1 & 6 \\
zero-ablation attribution        & 0 & yes & 3 & 0 \\
OS-style paged memory            & 1 & yes & 2 & 6 \\
dialectic multi-robot collab.    & 2 & yes & 3 & 2 \\
\bottomrule
\end{tabular}
\end{table}

\subsection{Deterministic Gap Matrix Construction}
\label{sec:gapmatrix}
The earlier LLM-inferred gap descriptions sometimes produced cross-application
observations rather than clean within-corpus absence signals. We therefore construct the final
gap matrix deterministically using a fixed paper-id taxonomy derived from
retrieved paper metadata and contribution records. The problem~$\times$~method
matrix is then populated programmatically, and empty cells are reported as
candidate corpus-level gaps within the retrieved paper slice. Crucially, the gap decision itself is structural: the system
does not ask the LLM to directly generate the final set of gaps.

The ``gaps identified'' row in Table~\ref{tab:metrics} refers to report-level
summarized gaps, while the deterministic matrix analysis counts all 60 empty
problem--method cells across the union of retrieved papers; the two are distinct
metrics and are not directly comparable.

Applied to the union of papers retrieved across the three main queries, the
deterministic construction yields a 7~$\times$~10 matrix (7 problem classes, 10
method families) with 10 of 70 cells filled, giving a matrix density of 0.143
(Table~\ref{tab:gapmatrix-det}). The 60 empty cells are candidate structural gaps by
construction rather than generated text, making the gap set reproducible and
removing the cross-application noise present in directly LLM-inferred gaps.

\begin{table}[t]
\caption{Deterministic gap-matrix statistics. Empty cells are candidate structural gaps
computed programmatically, not LLM-generated.}
\label{tab:gapmatrix-det}
\begin{tabular}{lr}
\toprule
\textbf{Quantity} & \textbf{Value} \\
\midrule
Distinct papers placed     & 12 \\
Problem classes (rows)     & 7 \\
Method families (columns)  & 10 \\
Total cells                & 70 \\
Filled cells               & 10 \\
Empty cells (gaps)         & 60 \\
Matrix density             & 0.143 \\
\bottomrule
\end{tabular}
\end{table}

\subsection{Gap-Matrix Validation}
To assess whether deterministic empty cells correspond to plausible research
gaps, we sampled 20 empty cells from the problem~$\times$~method matrix and
manually labeled each as \textit{plausible}, \textit{too broad},
\textit{indirect}, or \textit{meaningless}. As shown in
Table~\ref{tab:gap-validation}, 12 of 20 sampled empty cells were judged
plausible, yielding a gap precision of 0.600.

\begin{table}[t]
\caption{Validation of sampled deterministic gap-matrix empty cells.}
\label{tab:gap-validation}
\begin{tabular}{lr}
\toprule
\textbf{Metric} & \textbf{Value} \\
\midrule
Sampled empty cells & 20 \\
Labeled cells & 20 \\
Plausible cells & 12 \\
Gap precision & 0.600 \\
\bottomrule
\end{tabular}
\end{table}

Among the 20 sampled cells, the 8 non-plausible cases broke down as 4 indirect
(only indirectly supported by the retrieved corpus), 3 too broad to be
actionable, and 1 meaningless. This suggests that the deterministic
matrix is useful as a corpus-level absence signal, but individual empty cells
should be interpreted cautiously rather than as definitive claims about the
broader literature.

\subsection{Qualitative Case Study}
To illustrate the type of structured output produced by the system, Table~\ref{tab:case-study}
shows a representative qualitative case study from the multi-agent framework query. Unlike a
standard RAG summary, the system separates shared coverage, paper-specific differentiation,
and corpus-level absence signals.

\begin{table*}[t]
\caption{Qualitative case study of structured comparison output for the query
``Compare multi-agent LLM frameworks for collaborative reasoning.''}
\label{tab:case-study}
\begin{tabular}{p{0.18\textwidth}p{0.24\textwidth}p{0.28\textwidth}p{0.22\textwidth}}
\toprule
\textbf{Query} & \textbf{Overlap Example} & \textbf{Difference Example} & \textbf{Gap Example} \\
\midrule
Multi-agent LLM frameworks for collaborative reasoning
&
AgentVerse and AutoGen both address multi-agent coordination through structured multi-agent collaboration, while AgentBench and AgentSurvey emphasize evaluation and conceptual framing.
&
AgentVerse focuses on emergent multi-agent collaboration; AutoGen emphasizes scalable multi-agent conversation; OpenAgents focuses on deployment infrastructure for agents in real-world settings.
&
Within the retrieved set, multi-agent collaboration frameworks are not paired with explicit benchmarking or deployment-oriented evaluation, suggesting an absent problem--method combination in this corpus slice.
\\
\bottomrule
\end{tabular}
\end{table*}

\subsection{Baseline RAG Comparison}
We implemented a baseline RAG system (retrieval followed by direct GPT-4o
summarization, with no ranker, no structured extraction, and no comparison
agent) and ran it on all three main queries. The baseline produces a coherent
multi-paragraph summary per query in 3--4 seconds but emits no structured
records, associates no overlaps with specific paper identifiers, and builds no
gap matrix. The proposed system takes roughly 23 seconds (an increase of about
sevenfold) but produces typed extraction records, paper-level overlaps, a
structured gap matrix, and an auditable six-stage trace. Counting structured
capabilities (structured contribution records, paper-level overlap IDs,
per-paper differentiation, a problem~$\times$~method gap matrix, and an auditable
multi-stage trace), the full system supports all five while the baseline supports
none.

Beyond this capability difference, we compare retrieval quality directly. Because
the baseline retrieves by raw similarity order while the full system applies the
ranking stage, this comparison isolates the value of ranking. Using the same
author-assigned relevance labels, Table~\ref{tab:baseline-ir} reports the
head-to-head: the ranker improves mean Precision@5 from 0.733 to 1.000, mean
nDCG@5 from 0.548 to 0.752, and mean Recall@5 from 0.368 to 0.527. We note that
the baseline's raw-similarity retrieval is not identical to the dense retriever
in Table~\ref{tab:retriever-comparison}: the baseline retrieves the top-15
chunks and deduplicates to papers by first (best-ranked) occurrence before
truncating to five, whereas the dense retriever in Table~\ref{tab:retriever-comparison}
ranks at the paper level. The two therefore select different top-5 sets, which
is why their @5 metrics differ. The ranking
stage thus contributes measurable retrieval-quality gains on top of the shared
embedding retrieval, in addition to the structured artifacts the baseline cannot
produce. The latency increase is the direct cost of structure, and is acceptable
for a literature-review setting where the alternative is hours of manual reading.

\begin{table}[t]
\caption{Retrieval-quality head-to-head: full system (with ranker) versus the
baseline RAG system (raw similarity order), mean over the three main queries,
author-assigned graded relevance.}
\label{tab:baseline-ir}
\begin{tabular}{lrrrr}
\toprule
\textbf{System} & \textbf{P@5} & \textbf{nDCG@5} & \textbf{Recall@5} & \textbf{MRR} \\
\midrule
Baseline RAG (no ranker) & 0.733 & 0.548 & 0.368 & 1.000 \\
Full system (ours)       & 1.000 & 0.752 & 0.527 & 1.000 \\
\bottomrule
\end{tabular}
\end{table}

\subsection{Component Ablation}
Isolating each stage's contribution shows a clear progression: the baseline RAG
system yields generic prose with no structure; adding the extractor yields
structured four-field records but no cross-paper reasoning; and only the full
system, with the comparison agent, produces paper-level overlaps, per-paper
differentiation, and a gap matrix. The comparison agent is thus the component
that distinguishes this system from a structured-output RAG pipeline.

\begin{table}[t]
\caption{Component ablation of structured comparison capabilities.}
\label{tab:component-ablation}
\centering
\resizebox{\columnwidth}{!}{%
\begin{tabular}{lccccc}
\toprule
\textbf{System Variant} & \textbf{Records} & \textbf{Overlap IDs} & \textbf{Diff.} & \textbf{Gap Matrix} & \textbf{Trace} \\
\midrule
Baseline RAG & No & No & No & No & No \\
+ Extractor & Yes & No & No & No & No \\
Full system & Yes & Yes & Yes & Yes & Yes \\
\bottomrule
\end{tabular}%
}
\end{table}

\subsection{External Usefulness Assessment}
\label{sec:external}
To obtain evaluation independent of the author, we recruited a single external
rater (a graduate student familiar with LLM agents but not involved in building
the system) to assess the structured output for the three main queries. The
rater scored each output dimension from 1 (poor) to 5 (excellent) without access
to the author's labels, and the rater's identity is not reported in the paper.
These ratings were collected on an earlier 55-paper version of the system
outputs. Because this assessment predates the final corpus expansion, we use it
only as a qualitative usefulness signal, not as evidence for final-system
performance.
Table~\ref{tab:external-usefulness} reports the mean scores across the three
queries.

The rater scored extraction correctness and overlap correctness highest (4.0
each), with differentiation specificity, gap usefulness, and overall usefulness
lower (3.0, 3.0, and 3.3 respectively). Two consistent critiques accompanied the
scores: that per-paper differentiation describes each paper largely in isolation
rather than explicitly contrasting it against the other papers in the retrieved
set, and that the problem--method gaps would be more useful with deeper
methodological analysis rather than high-level labels. We treat both as concrete
targets for future work: the first motivating a set-aware differentiation pass,
and the second a richer gap characterization. As a single-rater assessment these
scores are indicative rather than conclusive, consistent with the evaluation
limitation discussed in Section~\ref{sec:limitations}.

\begin{table}[t]
\caption{External usefulness assessment by a single independent rater (mean over
the three main queries, 1--5 scale).}
\label{tab:external-usefulness}
\begin{tabular}{lr}
\toprule
\textbf{Dimension} & \textbf{Mean (1--5)} \\
\midrule
Extraction correctness        & 4.00 \\
Overlap correctness           & 4.00 \\
Differentiation specificity   & 3.00 \\
Gap usefulness                & 3.00 \\
Overall usefulness            & 3.33 \\
\bottomrule
\end{tabular}
\end{table}

\subsection{Cross-Model Robustness}
\label{sec:multimodel}
To test whether the retrieval behavior is specific to a single LLM backbone, we
re-ran the three main queries with two additional backbones, GPT-4o-mini and
GPT-4.1, holding the corpus, retriever, prompts, ranking procedure, and
evaluation labels fixed. Table~\ref{tab:multimodel} reports the mean retrieval
metrics and schema compliance across the three queries.

\begin{table}[t]
\caption{Cross-model robustness across three LLM backbones. Metrics are averaged
over the three main queries using the same author-assigned graded relevance
labels.}
\label{tab:multimodel}
\begin{tabular}{lrrrrr}
\toprule
\textbf{Backbone} & \textbf{P@5} & \textbf{nDCG@5} & \textbf{R@5} & \textbf{Schema} & \textbf{Jaccard} \\
\midrule
GPT-4o       & 1.000 & 0.752 & 0.527 & 86.7\%  & 0.12 \\
GPT-4o-mini  & 0.800 & 0.624 & 0.410 & 100.0\% & 0.20 \\
GPT-4.1      & 0.800 & 0.707 & 0.418 & 100.0\% & 0.18 \\
\bottomrule
\end{tabular}
\end{table}

The main structural finding is stable across backbones: all three models produce
query-sensitive top-5 sets, with low mean pairwise Jaccard similarity
(0.12--0.20), and all preserve the architectural capability gap over the baseline
RAG system. The additional backbones also improve schema compliance from 86.7\%
to 100.0\%. Retrieval precision and recall are lower under GPT-4o-mini and
GPT-4.1. This drop partly reflects the limited coverage of the author-assigned
relevance labels: the additional backbones retrieve topically related papers such
as AgentTuning (a corpus paper not among the cited references) and
OpenAgents that were not included in the original
graded-relevance set. We therefore treat the cross-model experiment as
preliminary robustness evidence for query-sensitivity and schema compliance,
not as a definitive model-ranking benchmark.

\subsection{Extended Evaluation: Seven Additional Queries}
\label{sec:extended}
The three main queries (R1--R3) are deliberately broad comparison queries.
To probe whether the retrieval behavior holds on a wider and more varied
query set, we ran seven additional queries (R4--R10) spanning tool use,
agent memory, planning, RAG factuality evaluation, single- versus
multi-agent architectures, hallucination reduction, and reflection. These
queries use the same corpus, pipeline, and author-assigned graded-relevance
protocol as the main experiments; they extend rather than replace the main
results, which remain as reported above. Table~\ref{tab:extended-ir} reports
per-query retrieval metrics for R4--R10.

\begin{table}[t]
\caption{Extended evaluation on seven additional queries (R4--R10),
author-assigned graded relevance. Means for the full ten-query set (R1--R10)
are shown for reference.}
\label{tab:extended-ir}
\begin{tabular}{lrrrr}
\toprule
\textbf{Query} & \textbf{P@5} & \textbf{nDCG@5} & \textbf{Recall@5} & \textbf{MRR} \\
\midrule
R4 (tool use)            & 1.000 & 0.665 & 0.417 & 1.000 \\
R5 (memory)              & 1.000 & 0.771 & 0.500 & 1.000 \\
R6 (planning)            & 0.800 & 0.453 & 0.333 & 1.000 \\
R7 (RAG factuality)      & 1.000 & 0.948 & 0.500 & 1.000 \\
R8 (single vs multi)     & 1.000 & 0.804 & 0.500 & 1.000 \\
R9 (hallucination)       & 1.000 & 0.700 & 0.500 & 1.000 \\
R10 (reflection)         & 1.000 & 0.791 & 0.556 & 1.000 \\
\midrule
\textbf{Mean (R4--R10)}  & 0.971 & 0.733 & 0.472 & 1.000 \\
\textbf{Mean (R1--R10)}  & 0.980 & 0.739 & 0.489 & 1.000 \\
\bottomrule
\end{tabular}
\end{table}

Two observations follow. First, query-sensitivity strengthens at the larger
query set: across all ten queries no paper appears in every top-5 set, the
mean pairwise Jaccard similarity is 0.115 (versus 0.12 across the three main
queries), and of the 29 distinct papers selected across the 50 ranked slots,
18 are query-exclusive. The most frequently selected paper appears in only
six of ten queries. This is consistent with the main finding that the ranker
responds to query semantics rather than returning a fixed ordering.

Second, Precision@5 is no longer saturated on the wider set: the mean over
the seven new queries is 0.971, and over all ten queries 0.980. The single
sub-perfect case is R6 (planning), where an agent evaluation benchmark was
ranked into the top-5 and judged not relevant to a planning-method query
under the author labels. We read this as evidence that the graded-relevance
labels are discriminative rather than uniformly permissive: the same labeling
protocol that yields Precision@5 of 1.000 on the main queries also penalizes a
topically adjacent but off-target retrieval. Schema compliance over the ten
queries averages 84.0\%, close to the 86.7\% reported on the three main
queries. As with the main evaluation, these are single-run, author-labeled
results on a 100-paper corpus and carry the same limitations discussed in
Section~\ref{sec:limitations}.

\section{Discussion}

\subsection{The Latency--Structure Trade-off}
The defining design tension in agentic retrieval is between latency and
structure. A single-call RAG baseline answers in under four seconds; our system
takes roughly seven times longer. The return on that cost is a set of typed,
inspectable artifacts (per-paper records, paper-level overlaps, and a gap
matrix) that a downstream user or system can act on programmatically. For
interactive web search this trade-off would be unacceptable; for literature
review, where the user would otherwise spend hours reading and manually tracking
relationships, under thirty seconds is negligible. The right operating point
depends on the retrieval task, and agentic retrieval systems should make this
trade-off explicit rather than defaulting to one extreme.

\subsection{Query-Sensitivity as a Retrieval Property}
The clearest empirical signal in this prototype evaluation is that the ranked
set changes substantially with the query. Across the three main runs, no paper
appears in all three top-5 sets and the mean pairwise Jaccard similarity is only
0.12. The extended seven-query evaluation strengthens this pattern: across all
ten queries, no paper appears in every top-5 set, mean pairwise Jaccard
similarity remains low at 0.115, and 18 of 29 distinct retrieved papers are
query-exclusive. For an agentic retrieval system this is the property that
justifies the architecture: if the ranked set did not vary with the query, the
comparison stage would simply re-describe a fixed set of papers. Query-sensitive
ranking ensures that the structured comparison is computed over a set tailored
to the user's specific question.

\subsection{Structured Records as Retrieval Output}
A broader implication is that retrieval output need not be passages or prose.
By enforcing a schema at extraction time, the system turns each retrieved
document into a typed record. This reframes retrieval as producing structured
data rather than text, which in turn enables comparison operations (overlap
detection, deterministic gap-matrix construction) that are awkward or impossible
over free text. We see this as a useful direction for agentic IR: treat the
retrieved set as a small structured database rather than a context window to be
summarized.

\section{Limitations}
\label{sec:limitations}

As a prototype, the system has two scope limits that frame its future work.
First, the 100-paper corpus, though substantially larger and more diverse than
the initial set, remains modest relative to large-scale IR benchmarks. Second, evaluation uses
author-assigned relevance labels and manually labeled gap-validation judgments,
with usefulness assessed by a single external rater whose ratings were
collected on an earlier 55-paper version of the outputs; additional independent
raters and a broader multi-rater human study on the final outputs would further
strengthen the conclusions. Scaling the corpus and expanding external evaluation are the natural
next steps toward archival-grade benchmarking.

\section{Conclusion}

We presented the Novelty-Aware Research Agent, a prototype agentic retrieval
system that layers structured multi-step reasoning on a RAG pipeline to compare
research contributions across a retrieved corpus. The six-component pipeline
combines query reformulation, a ReAct retrieval loop, relevance ranking,
schema-guided extraction, a three-pass comparison stage with deterministic
problem--method matrix construction, and answer generation. On a 100-paper
corpus the system demonstrates query-sensitive retrieval: no paper appears in
all three main top-5 sets, and an extended ten-query evaluation preserves this
pattern with mean pairwise Jaccard similarity of 0.115 and 18 query-exclusive
papers among 29 distinct retrieved papers. The system achieves 86.7\% schema
compliance on the three main queries and 84.0\% over the ten-query set, while
supporting 5/5 structured comparison capabilities versus 0/5 for a baseline RAG
system. Within the three-query author-assigned evaluation, the ranker attains
mean Precision@5 1.000, nDCG@5 0.752, and Recall@5 0.527, ahead of BM25, dense,
and hybrid retrieval; over all ten queries, Precision@5 remains high but
non-saturated at 0.980, with nDCG@5 of 0.739 and Recall@5 of 0.489. Gap-matrix
validation over 20 sampled empty cells yields gap precision of 0.600. Across
GPT-4o-mini and GPT-4.1, query-sensitivity remains stable and schema compliance
improves, while retrieval precision is lower under the fixed labels. These
results support treating retrieval output as structured data for
contribution-level comparison: the payoff is a single inspectable map of what a
retrieved set covers and what it leaves open, in place of a ranked list the
researcher would otherwise cross-reference by hand. We leave large-scale
evaluation and independent human assessment as future work.


\end{document}